\documentclass[aps,prb,twocolumn,superscriptaddress,showpacs,floatfix]{revtex4}

\usepackage{graphicx}
\begin{document}
\title{A lattice dynamical treatment for the total potential
energy of single-walled carbon nanotubes and its applications:
relaxed equilibrium structure, elastic properties, and vibrational
modes of ultra-narrow tubes}
\author{Jin-Wu Jiang}
\affiliation{Institute of Theoretical Physics, Chinese Academy of
Sciences, Beijing 100080, China }
\author{Hui Tang}
\affiliation{Institute of Theoretical Physics, Chinese Academy of
Sciences, Beijing 100080, China }
\author{Bing-Shen Wang}
\affiliation{State Key Laboratory of Semiconductor Superlattice
and Microstructure
\\ and
Institute of Semiconductor, Chinese Academy of Sciences, Beijing
100083, China\\}
\author{Zhao-Bin Su}
\affiliation{Institute of Theoretical Physics, Chinese Academy of
Sciences, Beijing 100080, China }
\affiliation{Center for Advanced
Study, Tsinghua University, Beijing 100084, China}
\begin{abstract}
In this paper, we proposed a lattice dynamic treatment for the
total potential energy for single-walled carbon nanotubes
(SWCNT's) which is, apart from a parameter for the non-linear
effects, extracted from the vibrational energy of the planar
graphene sheet. The energetics, elasticity and lattice dynamics
are treated in terms of the same set of force constants
independent of tube structures. Based upon the proposal, we
investigated systematically the relaxed lattice configuration for
narrow SWCNT's, the strain energy, the Young's modulus and Poisson
ratio, and the lattice vibrational properties respected to the
relaxed equilibrium tubule structure. Our calculated results for
various physical quantities are nicely in consistency with
existing experimental measurements. Particularly, we verified that
the relaxation effect brings the bond length longer and the
frequencies of various optical vibrational modes softer; Our
calculation provides an evidence that the Young's modulus of
armchair tube exceeds that of the planar graphene sheet, and the
large diameter limits of the Young's modulus and Poisson ratio are
in agreement with the experimental values of the graphite; The
calculated radial breathing modes for the ultra narrow tubes with
diameter range between 0.2 - 0.5~nm coincide the experimental
results and the existing {\it ab initio} calculations with
satisfaction; For narrow tubes of diameter 2~nm, the calculated
frequencies of optical modes in tubule tangential plane as well as
those of radial breathing modes are also in good agreement with
the experimental measurement. In addition, our calculation shows
that various physical quantities of relaxed SWCNT's can actually
be expanded in terms of the chiral angle defined for the
correspondent ideal SWCNT's.
\end{abstract}

 \maketitle

\section{Introduction}
After the discovery of the carbon nanotubes (CNT's) in
1991,\cite{Iijima} there have been several methods to prepare
CNT's with ultra small diameters. Previously, CNT's were
synthesized in a free space\cite{Dresselhaus2, Smalley} and the
quotient of narrow CNT's is quite low. Afterwards Z. K. Tang {\it
et~al.}\cite{Tang1, Tang2, Tang3, Tang4} initiated to grew
single-walled carbon nanotubes (SWCNT's) inside a channel of
zeolite templates. The SWCNT's prepared in this way have the
diameter as small as 0.42 $\pm$ 0.02~nm. Actually there are three
kinds of tubes with different chiralities in this diameter range:
$(5,0)$, $(4,2)$, and $(3,3)$. A fabrication process was further
developed to prepare a type of samples with $(5, 0)$ and $(3, 3)$
only\cite{Tang7} with a very narrow diameter distribution. More
recently, a stable CNT's with diameter 0.3~nm is found inside a
multi-walled carbon nanotube.\cite{Ando} As shown by the density
functional studies\cite{Ando} that this CNT might be interpreted
as the armchair CNT $(2,2)$ with a radial breathing mode (RBM) at
787~cm$^{-1}$. All these technological improvements stimulated
progressively experimental and theoretical studies on the ultra
narrow SWCNT's.\cite{White2, Damnjanovic7, Tang5, Tang6, Tang8,
Tang9, Tang10, Tang11}

Since the narrow CNT's possess biggish curvatures, the equilibrium
geometries would deviate from the ideal geometry, i.e., deduced
from a seamlessly rolling up planar graphitic lattice sheet
referring to a chiral vector
$\vec{R}=n_{1}\vec{a}_{1}+n_{2}\vec{a}_{2}$.\cite{White} Various
methods have been developed to determine the equilibrium geometry
of free SWCNT's.\cite{White2, Jackie, Daniel, Kurti, Popov3} Among
others, in Ref.~\onlinecite{White2}, the author calculated the
total energy for narrow SWCNT's with a first-principles,
all-electron, self-consistent local-density functional
band-structure method. The calculated total energy can be
parameterized in terms of five profitable parameters --i.e.,
generalized motif variables (GMV's)--with the line group symmetry
intimately retained.  The equilibrium geometry is obtained by an
optimization procedure with respect to these GMV's. In
Ref.~\onlinecite{Jackie}, a molecular mechanical model for the
effective total potential energy is proposed to calculate the
equilibrium structure and the strain energy of achiral SWCNT's
which is a quadratic form of lattice site displacements with
respect to  the planar graphite sheet. Apart from deviations of
three bond lengths and three bond angles, a pyramidalization angle
is introduced to describe the energy associated with the
curvature. The calculated results are consistent with existing
numerical results based on {\it ab initio} calculations.

Moreover, based upon the relaxed equilibrium geometry provided by
{\it ab initio} method,\cite{White2} Milo\u{s}evi\'{c} {\it et
al.}\cite{Damnjanovic7} applied the line-group symmetry-based
force constant method to calculate the vibrational modes of
ultra-narrow SWCNT's, in which the fitted force constants of
narrow SWCNT's have been adjusted with respect to the provided
cylindrical web geometry.

Besides, the elastic properties for carbon nanotubes have
stimulated a large number of experimental and theoretical
studies.\cite{Lieber} In particular, the Young's modulus is a
significant physical quantity which reflects the striction of
equilibrium structure in response to the impressed load. In
Ref.~\onlinecite{Wu}, the axis Young's modulus of the SWCNT is
measured as 1.2~Tpa by atomic force microscope technique. By
observing their freestanding room-temperature vibrations in a
transmission electron microscope,\cite{Treacy1} the Young's
modulus for 27 nanotubes in the diameter range 1.0 - 1.5~nm were
measured and the average Young's modulus is 1.25-0.35/+0.45
Tpa.\cite{Treacy2} An interesting fact was acknowledged that the
Young's modulus of certain nanotubes is even higher than that of
bulk graphite.

There have been a large number of theoretical calculations for the
Young's modulus and the Poisson ratio. The {\it ab initio}
method,\cite{Daniel, Lier, Zhou} tight-binding
approximation,\cite{Rubio, Molina, ZhouXin} and force constant
model\cite{Lu} gave the results of Young's modules with a range
from 1 to 1.24~Tpa, and Poisson ratio in a range 0.16 - 0.27. It
further shows that the calculated results vary slightly with
different radii and chiralities. However, in
Refs~\onlinecite{Popov2, Jackie2, Guo, Robertson}, the calculated
Young's modulus or the Poisson ratio are found to be chirality and
tube radius-dependent. In Ref~\onlinecite{Popov2}, the Poisson
ratio of armchair tubes has the value larger than that of planar
graphene sheet while for zigzag tubes smaller, and its planar
sheet limit is 0.21. In Refs~\onlinecite{Jackie2} and
\onlinecite{Guo}, the Poisson ratio decreases with increasing the
tube diameter with the large diameter limit as 0.16. Anyhow, to
our knowledge the resolution of the experimental measurements for
the Young's modulus and Poisson ratio is still not enough
satisfactory yet and the varies correspondent theoretical
calculations are even more diverse.

As we have seen in above, {\it to figure out the total potential
energy of SWCNT's as the functional of cylindrical lattice
configuration is the key issue underlying the investigation of the
equilibrium structure, strain energy, Young's modulus and lattice
dynamics of ultra-narrow tubes.}

In this paper, we propose that the total potential energy for
SWCNT's can be extracted from the vibrational energy of the planar
graphene sheet with five force constants. In which the bi-linear
forms of the lattice site displacements are accounted with the
equilibrium lattice configuration of the planar graphite as the
reference point, while one of the potential term for the twist
motion needs to be improved by including a quartic term to account
the non-linear effect. One of the advantages of our proposal is
that the equilibrium structure, the elasticity, and the lattice
dynamic properties for all kinds of SWCNT's are calculated in
terms of the same set of six (5+1) force constants, which
essentially reflects an unified microscopic mechanism founded on
the five forms of motion of the graphite-type cylindrical lattice
sheet. Based upon the proposal, we investigate systematically the
relaxed lattice configuration for narrow SWCNT's, the strain
energy, the Young's modulus and Poisson ratio, and the lattice
vibrational properties respected to the relaxed equilibrium tubule
structure. The calculated results are nicely in consistency with
the existing experimental measurements.

In particular, our calculation reveals a generic feature that the
relaxation effect exhibits itself in stretching the bond lengths
as well as softening the mode frequencies for all kinds of
SWCNT's; It provides a kind of evidence that the Young's modulus
of armchair tubes exceeds that of the planar graphene sheet, i.e.,
when the radius increases, the calculated values of the Young's
modulus for armchair tubes approach that of the graphite from
above and those of zigzag tubes from below. Meanwhile, the large
diameter limits of the Young's modulus and Poisson ratio are in
agreement with the experimental values of the graphite; The
calculated RBM for the ultra narrow tubes with diameter range
between 0.2 - 0.5~nm coincide the experimental results\cite{Tang7}
and the existing {\it ab initio} calculations\cite{Ando} with
satisfaction. For narrow tubes of diameter 2~nm, the calculated
frequencies of optical modes in tubule tangential plane as well as
those of RBM are also in good agreement with the experimental
measurement;\cite{Meyer2} The honeycomb lattice symmetry-based
chiral expansion for physical quantities of SWCNT's with ideal
geometry\cite{Jiang} can be retained for those of relaxed SWCNT's
by changing the chiral angle into its counterpart defined in tubes
of ideal geometry; As expected, for tubes with diameter larger
than approximately 0.8~nm where the relaxation effect dies away,
the calculated frequencies and the sound velocities for various
vibration modes are in accord with those of the tubes of ideal
geometry.\cite{Jiang}

In Section II, the  total potential energy for SWCNT's is
presented, and the relaxed equilibrium geometry and correspondent
strain energy are discussed. Section III is devoted  to the
lattice dynamics for ultra-narrow SWCNT's, while Section IV for
the Young's modulus and Poisson ratio. Finally, the conclusion is
in Section V.
\section{The relaxed equilibrium structure}
\subsection{The geometry}
\begin{figure}
    \begin{center}
        \scalebox{0.8}[0.8]{\includegraphics[width=7cm]{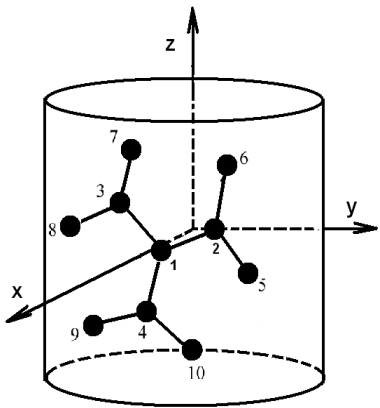}}
    \end{center}
    \caption{Sketch of carbon atoms on a cylindric surface.}
    \label{Fig:nanotube}
\end{figure}

We recall the geometrical description for the relaxed equilibrium
structure of SWCNT's particularly for the ultra-narrow tubes
following Ref.~\onlinecite{White2}. As for the cylindrical
coordinate description shown in Fig.~\ref{Fig:nanotube}, we first
set the $z$ axis along the tube axis, while the $x$ axis is fixed
passing through one of the $A$ atoms. We further introduce a
planar image of the cylindrical lattice sheet by unfolding the
cylindrical tube into a planar lattice sheet, which is actually a
tangential plane with the tangent line as a generatrix passing
through the $A$ atom, and set atom $A$ as the origin of the plane
with coordinates as $(r_A,0,0)$, where $r_A$ is the radius of atom
$A$. The planar lattice sheet is constituted of parallelogram
primitive cells with $\vec{a}_1$ and $\vec{a}_2$ as its basic
vectors. The deviations of CNT's with relaxed geometry from that
of ideal geometry can be illustrated most conveniently by the
primitive cells on the image plane. For that of ideal geometry,
the basic vectors are of equal magnitude with an intra-angle of
$\pi / 3$ between themselves, while for that of the relaxed
geometry, their magnitudes are no more equal and the angle can be
deviated from $\pi /3$. Meanwhile, the location of the in-cell $B$
atom relative to $A$ atom could also be deviated from that of the
ideal geometry.

Now we introduce the chiral vector
$\vec{R}=n_{1}\vec{a}_{1}+n_{2}\vec{a}_{2}$ initiated from the
atom $A$ along the $y$ axis in the image sheet, and $R$ is the
image circumference of cylindrical lattice sheet. We then
introduce a screw vector $\vec{H}=p_1\vec{a}_1+p_2\vec{a}_2$ with
integers $p_1$ and $p_2$ satisfying $p_1n_2-p_2n_1=N$ and $N$ is
the largest common divisor of the integer pair $(n_1,n_2)$.
Exactly following Refs~\onlinecite{White2} and \onlinecite{White},
$\vec{H}$ and $\frac{\vec{R}}{N}$ constitute the basic ingredients
of the generalized motif cell, upon which a generalized screw
operation $S(\alpha, h)$ and a generalized $N$-fold rotation
operation $C_N$ can be established. The former is a rotation
$\alpha$ around the tube axis with a simultaneous translation $h$
along the tube axis, the latter a rotation of $\frac{2\pi}{N}$
around the tube axis. A generalized motif cell $(m,l)$ can be
generated from the $(0,0)$ cell  by successive operations of
$S(\alpha,h)$ and $C_N$ for $m$ and $l$ times respectively. For
the in-cell geometry of a generalized motif cell, originated from
atom $A$ the position of atom $B$ can be fixed by a rotation
$\alpha'$ around the tube axis and two simultaneous translations
$h'$ and $(r-r_{A})$ along the tube axis and the tube radius
respectively.

As a result, the coordinates on the cylindrical lattice sheet for
each site $(m,l,s)$ with $s=A,B$ can be expressed as
\begin{eqnarray}
 &  \vec{r}(m,l,s)=(r_{s}\cos\phi, r_{s}\sin\phi, z)\; , \\
 &     r_{s}=r_{A}, \hspace{0.25cm}
     \phi=m\alpha+\frac{l}{N}2\pi, \hspace{0.25cm}
     z=mh \;\;\; (\mbox{for $A$}),\nonumber \\
 &     r_{s}=r, \hspace{0.25cm}
     \phi=m\alpha+\frac{l}{N}2\pi r+\alpha', \hspace{0.25cm}
     z=mh+h' \;\;\; (\mbox{for $B$}).\nonumber
\end{eqnarray}
While the corresponding coordinates in the image plane have the
form as
\begin{eqnarray}
   \vec{r}^{\;{\scriptsize\mbox{image}}}(m,l,s)=(r_{s}, \phi r_{s}, z)
\end{eqnarray}
with the $x$-components of both $A$ and $B$ atoms as constants
respectively.

Therefore, the six GMV's $r$, $r_{A}$, $\alpha'$, $h'$, $\alpha$,
and $h$ provide a complete description for the geometry of the
lattice structure of SWCNT's, in which all the $A$ and $B$ atoms
are allowed to  sit on different cylindrical surfaces with radii
$r_A$ and $r$ respectively.
\subsection{The total potential energy}
In fact, in sense of quadratic approximation the vibration
potential of the single layer graphite lattice sheet provides a
proper description of various modalities of lattice motion. We,
therefore, propose that the total potential energy of SWCNT's can
be extracted from the vibration potential of graphene lattice
sheet\cite{Aizawa} as  follows. Firstly, the total potential
energy contributed by carbon atom 1, i.e., atom $(0,0,A)$, is
composed of the following five terms in which a local planar
graphene sheet is taken as the reference point,
\begin{eqnarray}
    U_{A}=V_{l}+V_{sl}+V_{BB}+V_{rc}+V_{tw} \; ,
    \label{EnergyA}
\end{eqnarray}
where $V_{l}$, $V_{sl}$ are the potentials of the spring forces
between the nearest-neighbor and the next-nearest-neighbor atom
pair respectively as shown in Fig.~\ref{Fig:nanotube}, $V_{BB}$ is
the energy associated with bond angle variations, $V_{rc}$
describes out-of-surface bond bending, i.e., a kind of strain
force on atom $i$ in out-of-surface direction  by its three
nearest neighbors in curling process, and $V_{tw}$ is the twist
potential energy.
\begin{eqnarray}
V_{l} & = & \frac{k_{l}}{4}\sum_{i=2}^{4}
    [(\Delta\vec{r}_{i}-\Delta\vec{r}_{1})\cdot\vec{e}_{1i}^{\;l}]^{2}\;,
    \label{Potential1}\\
V_{sl} & = & \frac{k_{sl}}{4}\sum_{j=5}^{10}
    [(\Delta\vec{r}_{j}-\Delta\vec{r}_{1})\cdot\vec{e}_{1j}^{\;l}]^{2}\;,
    \label{Potential2}\\
V_{BB} & = & \frac{k_{BB}}{2}\sum_{\eta=1}^{3}
    (\cos\Theta_{\eta}-\cos\Theta_{0})^{2}
    \; ,    \label{Potential3}\\
V_{rc} & = &
\frac{k_{rc}}{2}\left[\left(\sum_{i=2}^{4}\Delta\vec{r}_{i}-3\Delta\vec{r}_{1}\right)
    \cdot\vec{e}_{1}^{\;r}\right]^{2},
    \label{Potential4}\\
V_{tw} & = & \frac{k_{tw}}{4}\sum_{\left<i,j\right>}f(x_{ij}^1)\; ,
    \label{Potential5}\\
x_{ij}^1 & = & [(\Delta\vec{r}_{i}-\Delta\vec{r}_{j})
    -(\Delta\vec{r}_{i'}-\Delta\vec{r}_{j'})]
    \cdot\vec{e}_{1k}^{\;r} \; , \nonumber \\
f(x) & = & \left\{
\begin{array}{cc}
x^{2}-K_{ah}x^{4}, & \mbox{if} \hspace{0.4cm} |x|<\sqrt{\frac{1}{2K_{ah}}}\; ,\\
\frac{1}{4K_{ah}},& \mbox{if} \hspace{0.4cm}
|x|\geq\sqrt{\frac{1}{2K_{ah}}}\; .
\end{array}
\right.  \nonumber
\end{eqnarray}
In Eqs~(\ref{Potential1}) to (\ref{Potential5}) $i=2,\ldots ,4$
and $j=5,\ldots, 10$ are the nearest and next-nearest neighbors of
atom 1 respectively, $\Theta_{\eta}$ with $\eta=1,2,3$ represent
the three bond angles with atom 1 as the common apex while
$\Theta_{0}=\frac{2\pi}{3}$ is the bond angle in the graphic
plane, $\left< i,j\right>$ represents a pair of atoms nearest
neighbored to atom 1 with $k$ being the third of its nearest
neighbors and the pair $\left< i',j'\right>$ is the image of
$\left< i,j\right>$ referring to a $C_2$ rotation around the axis
in $\vec{e}_{1k}^{\; r}$. Moreover,
\begin{eqnarray}
   \Delta\vec{r}_i=\vec{r}_i-\vec{r}^{\;0}_{i}
\end{eqnarray}
is the displacement of the carbon atom $i$ from the planar
graphene sheet $\vec{r}^{\;0}_i$ to its counterpart in the tubal
lattice sheet $\vec{r}_i$, in which $i$ stands the site index
$(m,l,s)$ defined on the cylindrical lattice sheet. In addition,
$\vec{e}^{\; l}_{1i}=\frac{\vec{r}^{\;0}_i
-\vec{r}^{\;0}_1}{|\vec{r}^{\;0}_i-\vec{r}^{\;0}_1|}$ is the unit
vector pointed from atom 1 to atom $i$ in the graphite sheet,
$\vec{e}^{\; r}_{i}$ and $\vec{e}^{\; r}_{1k}$ are vertical unit
vectors of the planar graphene sheet located at the site $i$ and
the middle point of sites 1 and  $k$ of the graphene sheet
respectively. All the unit vectors are defined on the local
reference graphene sheet and are introduced to keep the rigid
rotational invariance symmetry. The above five potential terms  in
principle cover the main features of the modalities of lattice
deformation suggested by Lianxi Shen {\it et~al.},\cite{Jackie,
Jackie2} Chang {\it et~al.}\cite{Guo} in their mechanical model.

Correspondingly, that part of total potential energy in
association with carbon atom $B$ $(0,0,B)$ has a similar
expression as that of $U_{A}$, but with its geometrical parameters
replaced by the counterpart of atom $B$.

Secondly, we express the lattice displacements in $U_A$ and $U_B$
in terms of the generalized motif variables. It spells that the
locally constructed $(0,0)$ generalized motif cell with a pair of
carbon atoms $A$ and $B$ can be continued to form a whole
seamlessly closed cylindrical lattice sheet in which the
generalized screw as well as rotation symmetries are inherited. As
a result the total potential energy of SWCNT's has the form as a
simple multiple of those associated with a pair of nearest
neighboring atoms $A$ and $B$, i.e., $U_{A}$ and $U_{B}$
respectively,
\begin{eqnarray}
U=\mathcal{N}\ast (U_A+U_B) \; ,
\end{eqnarray}
where $\mathcal{N}$ is the number of generalized motif cells of
SWCNT's in consideration.  The redundant counting in the total
potential energy contributed by different carbon atoms can be
accounted by a proper resetting for the values of the relevant
force constants. We further address that this sort of
parameterization procedure can be carried out for all kinds of
SWCNT's including both chiral and achiral tubes.

The values of force constants are $k_{l}=305$~N m$^{-1}$,
$k_{sl}=68.25$~N m$^{-1}$, $k_{BB}=1.38\times10^{-11}$~erg,
$k_{rc}=14.8$~N m$^{-1}$, $k_{tw}=6.24$~N m$^{-1}$, and
$K_{ah}=2.5$. The frontal five force constants come from vibration
energy of planar graphene sheet\cite{Aizawa}. Since the twisting
deformations increase substantially with decreasing the tube
diameters,  for example the twisting angle for the C-C bond can
take the value as large as $26^{\circ}$ for the narrow SWCNT
$(6,3)$, an additional force constant for the anharmonic
improvement as $K_{ah}$ in Eq.~(\ref{Potential5}) has to be
introduced. We notice here that the differences in lattice
dynamics between such two types of systems are essentially due to
their different spatial geometry, which are characterized by their
own equilibrium lattice structure.
\subsection{The relaxed structure of SWCNT's}
We introduce the total potential energy per atom as $E$,
\begin{equation}
E=E(r,r_{\sl A},\alpha,h,\alpha',h')=(U_{A}+U_{B})/2 \; .
\end{equation}
By minimizing $E(r,r_{\sl A},\alpha,h,\alpha',h')$  with respect
to the six GMV's, we obtain the optimized equilibrium geometry
described by GMV's $(\bar{r}, \bar{r_{A}}, \bar{\alpha}, \bar{h},
\bar{\alpha'}, \bar{h'})$ as well as the strain energy
$E_s=E(\bar{r}, \bar{r_{A}}, \bar{\alpha}, \bar{h}, \bar{\alpha'},
\bar{h'})$. It is interesting to address that the optimized
equilibrium lattice configurations for all tubes are shown to be
of $\bar{r_A}=\bar{r}$. This is because that the total energy $U$
keeps unchanged when the tube is rotated upside down with atoms
$A$ and $B$ mutually permuted, so that the optimized geometry must
exhibit the $C_2$ symmetry. The reflect symmetry $\sigma_{h}$ with
respect to the cross section is also kept for achiral tubes.

\begin{table}[t]
\caption{The calculating results of the five GMV's with
         $\bar{r_A}=\bar{r}$ and strain
         energy for narrow SWCNT's.} \label{Tab:geometry}
\begin{ruledtabular}
\begin{tabular}{crrrrrr}
$(n_{1}, n_{2})$ & $r$~(\AA) & $\alpha'$~(rad) & $h'$~(\AA) &
$\alpha$~(rad)   & $h$~(\AA) & $E_{s}$~(eV)               \\
\hline
(2, 2) & 1.743 & 1.1036 & 0.000 & 1.5708 & 1.255 & 1.5596 \\
(3, 1) & 1.817 & 1.0019 & 0.337 & 4.5923 & 0.580 & 1.7596 \\
(4, 0) & 1.977 & 0.7854 & 0.613 & 0.7854 & 2.033 & 1.3636 \\
(3, 2) & 2.013 & 0.8460 & 0.148 & 2.4777 & 0.491 & 0.7977 \\
(4, 1) & 2.129 & 0.7512 & 0.416 & 4.9317 & 0.455 & 0.8171 \\
(3, 3) & 2.281 & 0.7088 & 0.000 & 1.0472 & 1.238 & 0.4522 \\
(5, 0) & 2.283 & 0.6283 & 0.652 & 0.6283 & 2.069 & 0.6484 \\
(4, 2) & 2.333 & 0.6788 & 0.250 & 2.0127 & 0.803 & 0.4781 \\
(5, 1) & 2.458 & 0.6075 & 0.475 & 5.1657 & 0.376 & 0.4606 \\
(4, 3) & 2.593 & 0.6000 & 0.112 & 1.7827 & 0.351 & 0.3083 \\
(6, 0) & 2.617 & 0.5236 & 0.672 & 0.5236 & 2.089 & 0.3759 \\
(5, 2) & 2.673 & 0.5652 & 0.324 & 2.6560 & 0.339 & 0.3163 \\
\end{tabular}
\end{ruledtabular}
\end{table}
\begin{table}[t]
\caption{Contributions of five potential energy terms to the
         strain energy for tube $(3,3)$.}\label{Tab:dominate}
\label{Tab:EnergyTerms}
\begin{ruledtabular}
\begin{tabular}{cccccc}
(3, 3) & $V_{l}$ & $V_{sl}$ & $V_{BB}$ & $V_{rc}$ & $V_{tw}$\\
\hline
(eV) & 0.02934 & 0.07191 & 0.02972 & 0.30179 & 0.01948\\
\end{tabular}
\end{ruledtabular}
\end{table}

\begin{figure}[htpb]
    \begin{center}
        \scalebox{0.8}[0.8]{\includegraphics{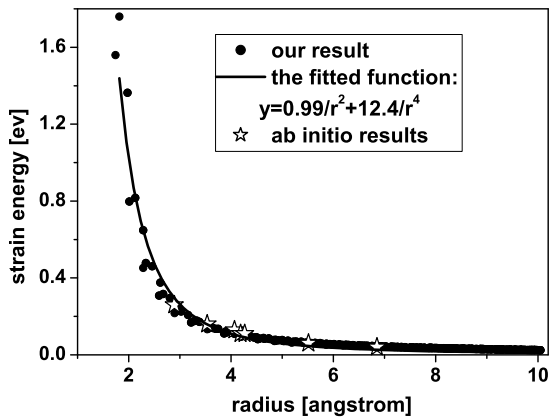}}
    \end{center}
    \caption{The strain energy as the function of radius for
    all tubes in $[1.3, 10.0]\,${\AA}. It is compared with the
    {\it ab initio} calculations.\cite{Jackie, Daniel}}
    \label{Fig:strainenergy}
\end{figure}

The calculated various lattice structure properties and the strain
energies for narrow SWCNT's with diameter in [0.26, 0.5]~nm are
listed in Table~\ref{Tab:geometry}, where the calculated GMV's are
in good agreement with those by \textsl{ab initio}
calculations.\cite{White2}  Moreover, it is shown that the
curvature energy associated with the fourth potential energy term
in Eq.~(\ref{Potential4}) dominates the total strain energy, while
the corresponding contributions from the bond variation are
usually much less than those from the bond angle variations (see
Table~\ref{Tab:dominate}).

In Ref.~\onlinecite{Robertson}, it is reported that the strain
energy various with tubule radius as $\frac{1}{r^{2}}$ by applying
the Tersoff-Brenner interatomic potentials. We here plot the
strain energy versus radius in Fig.~\ref{Fig:strainenergy} which
is comparable with the existing result\cite{Daniel, Robertson} and
can be well fitted by
$E_{s}=\frac{c_{1}}{r^{2}}+\frac{c_{2}}{r^{4}}$ in consistency
with the result of Ref.~\onlinecite{Jackie}. The coefficient of
$\frac{1}{r^{4}}$--i.e., $c_{2}$--is a bit large. That is because,
in Ref.~\onlinecite{Jackie}, there is no such twisting energy term
as we have.
\begin{figure}[htpb]
    \begin{center}
        \scalebox{1.5}[1.5]{\includegraphics{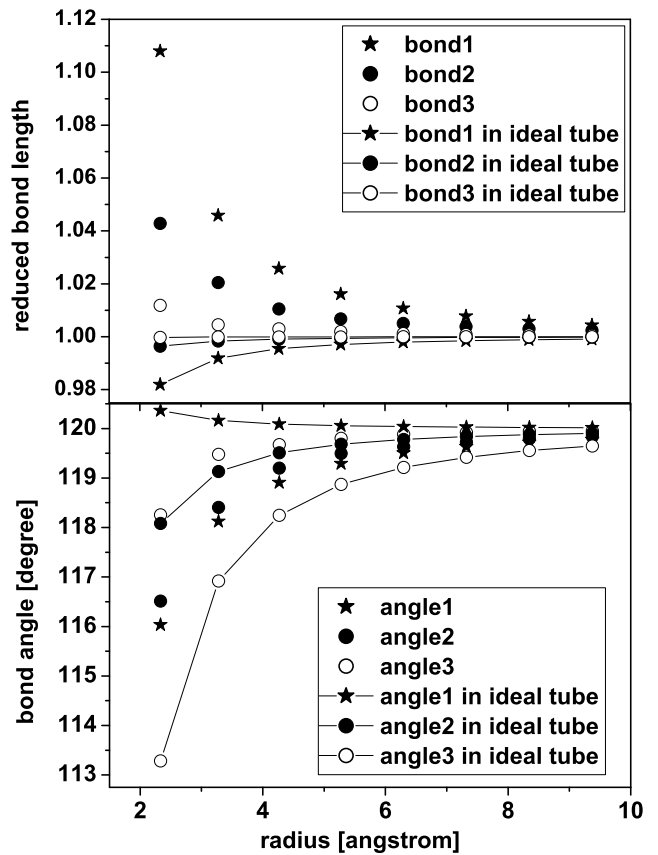}}
    \end{center}
    \caption{C-C bond lengths reduced by that of graphite and bond
    angles for different chiral tubes $(2n, n)$ with the relaxed geometry
    and ideal geometry.}
    \label{Fig:bondlength}
\end{figure}

Our calculation shows that the radius and C-C bond lengths of
CNT's of the optimized lattice configuration are always larger
than those of the correspondent CNT's of the ideal geometry. We
representatively display the three bond lengths in unit of the
graphene bond length $0.142$~nm for $(2n,n)$ tubes with the
relaxed geometry as well as ideal geometry in
Fig.~\ref{Fig:bondlength}a. The relaxation, resulted from the
optimization procedure, would disentangle the warping-induced
tension and make the inter atomic bond length getting longer. To
be precise, the bond lengths of the optimized lattice
configuration decrease with increasing diameter and keep their
values always larger than that of the graphene sheet. In contrary,
for tubes of ideal geometry the bond lengths increase with
increasing diameter and keep their values always smaller than that
of the graphene sheet. As we can see in
Fig.~\ref{Fig:bondlength}a, one of the bond lengths of the tube
$(4,2)$ is $0.139$~nm for ideal geometry while $0.157$~nm for case
of relaxed geometry which is about $11\%$ more than that of the
graphene sheet.

As shown in Fig.~\ref{Fig:bondlength}b, deviations of the bond
angles from that of the graphene  are really considerable for
narrow tubes with diameter smaller than $0.8$~nm. On the other
hand, as can be seen in the same figure, this effect decreases
quickly with increasing the tube diameter.

\section{Calculation results and discussion of phonon dispersion}
In Sec. II, the relaxed equilibrium lattice configurations of
SWCNT's are derived by an optimization procedure applied to the
proposed total potential energy. The vibrational potential of
SWCNT's is actually caused by the deviation of the lattice sites
with respect to the relaxed equilibrium lattice configuration. It
is obvious that the vibrational potential of SWCNT's would take
the same form as the total potential energy, whereas the
equilibrium position of the lattice site in the planar graphene
sheet is therefore replaced by those in the relaxed cylindrical
lattice sheet, meanwhile, the non-linear term is no more needed.
The detailed modifications are as follows.

(1) $\Delta \vec{r}_i=\vec{r}_i-\vec{r}_{i0}$ is renamed  by
convention as the vibrational displacement $\vec{u}_i$  with
$\vec{r}_{i0}$ the local planar graphene sheet to be replaced by
the relaxed equilibrium position of atom $i$ in SWCNT's;

(2) $ \cos \Theta _{0}$ in Eq.~(\ref{Potential3}) is no more the
bond angle of graphene sheet as $\frac{2}{3} \pi$ but the
correspondent bond angle of the relaxed equilibrium configuration
which depends on distinct adjacent atom pairs nearest neighbored
to the common apex atom $i$;

(3)  $\vec{e}_{1i}^{\;l}$ and $\vec{e}_{1j}^{\; l}$ are kept to be
the unit vectors pointed from atom $i$  and $j$ to atom 1
respectively. $\vec{e}_{1}^{\; r}$ is replaced by $\vec{e}_1^{\;
rc}=-\frac{\sum_{i=2}^4 \vec{r}_{i0}}{| \sum_{i=2}^4 \vec{r}_{i0}
|}$ and $\vec{e}_{1k}^{\; r}$ becomes now  the unit vector along
the radial direction of the middle point of atoms 1 and $k$. We
stress that all these unit vectors are defined on the cylindrical
lattice sheet with optimized equilibrium geometry, which {\it
taken a crucial role} to keep the rigid rotational invariance for
the vibrational potential of relaxed SWCNT's;

(4) The coefficient $K_{ah}$ is taken to be zero.

We emphase here that {\it the five force constants of the above
proposed quadratic vibrational energy are the same as those in
~Sec IIB. They are applied to all kinds of SWCNT including those
with small radii.} This is because that the rigid rotation
symmetry of the SWCNTs is precisely kept. And the curvature effect
for the vibrational modes has been properly taken care by the
above introduced quadratic expression.

It is straightforward that such obtained vibrational potential can
be again parameterized in terms of GMV's. Taking the advantage of
GMV description for the cylindrical lattice configuration, the
underlined generalized screw and rotation symmetries ensure that
the lattice dynamic equation even for tubes with relaxed geometry
can be decomposed into a six-dimensional eigenvalue problem and
becomes treatable.

We calculate, for tubes with diameter from 0.28~nm to 2.5~nm, all
the Raman- and Infrared-active modes,\cite{Alon} and the
velocities for the twisting modes (TW) and longitude acoustic
modes (LA).
\begin{figure}[htpb]
    \begin{center}
        \scalebox{0.8}[0.8]{\includegraphics{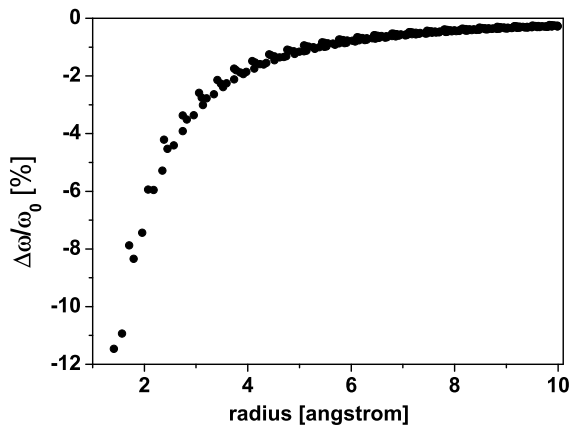}}
    \end{center}
    \caption{The radius dependence of relative frequency change
    due to the relaxation effect for the optical mode along the $z$ axis.
    The small waviness of the line shows the dependency of the
    frequencies on the chiral angle. For chiral and zigzag tubes
    this mode is the Raman-active mode.}
    \label{Fig:ModeVariety}
\end{figure}

\begin{table}[t]
\caption{The values of two optical modes in the tangential
         plane from our theory and compared with
         the experimental results.\cite{Meyer2}} \label{Tab:Optical}
\begin{ruledtabular}
\begin{tabular}{ccccc}
Optical mode & (11, 10) & (17, 9) & (27, 4) & (15, 14)  \\
\hline
Our theory & 1597.3 & 1600.5 & 1602.5 & 1602.1\\
           & 1600.7 & 1603.6 & 1605.5 & 1603.7\\
Experiment\cite{Meyer2} & 1566(7)   & 1572(13) & 1577.5(10) & 1573(7)\\
                        & 1593.5(6) & 1591(8)  & 1593(5)    & 1592(5)\\
\end{tabular}
\end{ruledtabular}
\end{table}

\begin{table}[t]
\caption{The values of RBM from our theory and compared with
         the experiments or other calculations. } \label{Tab:RBM}
\begin{ruledtabular}
\begin{tabular}{ccccc}
RBM & (2, 2) & (5, 0) & (3, 3) & (4, 2) \\
Our theory & 804.3 & 561.3 & 550.8 & 538.6 \\
comparison & $(787)^{a}$ & $(550)^{b}$ & / & $(510)^{b}$ \\
\hline RBM & & (11, 10) & (16, 7) & (15, 6) \\
Our theory &  & 160.0 & 142.5 & 155.2\\
comparison &  & $(169.5(7))^{c}$ & $(154(5))^{c}$ & $(166(1))^{c}$\\
\end{tabular}
\end{ruledtabular}
$^{a}$The {\it ab initio} calculations from
Ref.~\onlinecite{Ando}; $^{b}$The experiment results from
Ref.~\onlinecite{Tang7}; $^{c}$The experiment results from
Refs~\onlinecite{Meyer2} and ~\onlinecite{Meyer}.
\end{table}

\begin{figure}[htpb]
    \begin{center}
        \scalebox{0.8}[0.8]{\includegraphics{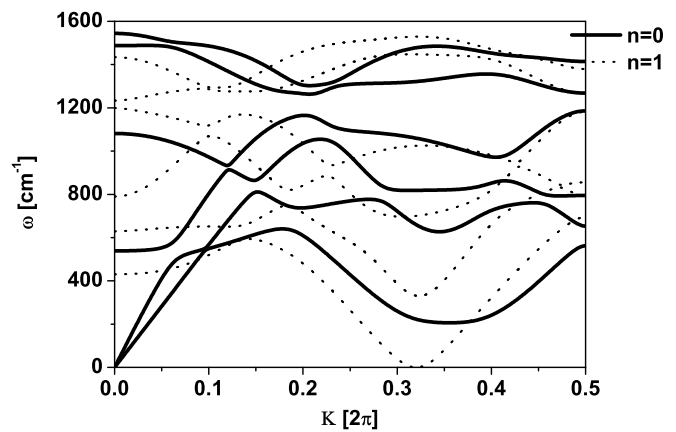}}
    \end{center}
    \caption{Phonon dispersion curves for SWCNT (4,2) in the
             representation of helical and rotational quantum numbers.
             The flexure mode---i.e., the transversal acoustic
             mode---shows up at $(\kappa, n)=(\alpha, 1)$, where $\alpha$ is
             in the unit of $2\pi$. It is double degenerate.}
    \label{Fig:SWCNT42}
\end{figure}

\begin{table*}[t]
\caption{Polarization vectors
$\vec{u}\equiv(\vec{u}(A),\vec{u}(B))$ as functions of $r$ ($r \in
[1.3, 10.0]$~\AA) and $\theta$. Where $\vec{u}(A)$ and
$\vec{u}(B)$ indicate the displacement vectors of  atoms $A$ and
$B$ in the $(0, 0)$ unit cell respectively. For the three modes in
this table, $u_{r}(B)=u_{r}(A)$, $u_{\phi}(B)=-u_{\phi}(A)$,
$u_{z}(B)=-u_{z}(A)$. } \label{Tab:eigenvector}
\begin{ruledtabular}
\begin{tabular}{ccll}
$(\kappa, n)=(0, 0)$ & & Vector($\theta$) & $f_{i}(r)$  \\
\hline
$R_{1}$ ($\vec{e}_{r}$ AC) & $u_{r}(A)$ & $f_{0}(r)$ & $f_{0}(r)=0.7072-\frac{0.0026}{r^{2}}$ \\
& $u_{\phi}(A)$ & $f_{1}(r)\sin3\theta$ & $f_{1}(r)=-\frac{0.0212}{r}+\frac{0.1111}{r^{2}}$ \\
& $u_{z}(A)$ & $f_{1}(r)\cos3\theta$ & $f_{1}(r)=-\frac{0.0344}{r}+\frac{0.2167}{r^{2}}$ \\
\hline
$R_{2}$ ($\vec{e}_{\phi}$ OP) & $u_{r}(A)$  & $f_{1}(r)\sin3\theta$ & $f_{1}(r)=\frac{0.0211}{r}-\frac{0.1064}{r^{2}}$ \\
& $u_{\phi}(A)$ & $f_{0}(r)+f_{1}(r)\cos12\theta$ & $f_{0}(r)=0.6975+\frac{0.0313}{r^{2}}$, $f_{1}(r)=0.0084-\frac{0.0248}{r^{2}}$\\
& $u_{z}(A)$ & $f_{1}(r)\sin6\theta+f_{2}(r)\sin12\theta$ & $f_{1}(r)=0.1642-\frac{0.3742}{r^{2}}$, $f_{2}(r)=-0.0283+\frac{0.1512}{r^{2}}$ \\
\hline
$R_{3}$ ($\vec{e}_{z}$ OP) & $u_{r}(A)$ & $f_{1}(r)\cos3\theta$ & $f_{1}(r)=\frac{0.0348}{r}-\frac{0.2189}{r^{2}}$\\
& $u_{\phi}(A)$ & $f_{1}(r)\sin6\theta+f_{2}(r)\sin12\theta$ & $f_{1}(r)=0.1643-\frac{0.3759}{r^{2}}$, $f_{2}(r)=-0.0283+\frac{0.1510}{r^{2}}$\\
& $u_{z}(A)$ & $f_{0}(r)+f_{1}(r)\cos12\theta$ & $f_{0}(r)=0.6975+\frac{0.0299}{r^{2}}$, $f_{1}(r)=0.0084-\frac{0.0236}{r^{2}}$\\
\end{tabular}
\end{ruledtabular}
\end{table*}

The relaxation effect for the non-zero vibrational modes of the
tubes with relaxed geometry exhibits itself as the frequency
lowering (mode softening) in comparison with their counterpart of
the tubes of ideal geometry. The effect of the relaxation on the
optical mode along the $z$ axis is shown in
Fig.~\ref{Fig:ModeVariety}. The relaxation softens this mode
considerably for narrow tubes. For example for the tubes with
radius as narrow as 2~{\AA}, this mode is softened as much as
$8\%$ in reference to that of the ideal geometry.

The calculated frequencies for the optical mode in the tubal
tangential plane coincide excellently with the experimental
measurements on tubes $(11,10)$, $(17,9)$, $(27,4)$,
$(15,14)$,\cite{Meyer2} see Table~\ref{Tab:Optical}. For the RBM,
the calculated results from our model are in good agreement with
the correspondent results from the experiments on tubes $(5,0)$,
$(4,2)$, $(11,10)$, $(16,7)$, $(15,6)$ listed in
Table~\ref{Tab:RBM}. Also for tube $(2,2)$ the calculate result is
consistent with that of the {\it ab initio} calculation\cite{Ando}
(see Table~\ref{Tab:RBM}). It can be seen moreover from
Table~\ref{Tab:RBM} that the smaller tube has the larger RBM
frequency. This is consistent with some other calculation
results.\cite{Damnjanovic7, Tang7}

We take SWCNT $(4,2)$ as a typical example of the narrow tubes.
Its dispersion is shown in Fig.~\ref{Fig:SWCNT42} with the helical
and rotational quantum numbers $\kappa$ and $n$ as $n=0,1$ and
$\kappa\in[0, \pi]$. In the neighborhood of $(\kappa, n)=(0, 0)$,
there are two zero-frequency modes, TW and LA, with acoustic
velocities as $C_{TW}=13.6$~km~s$^{-1}$, and
$C_{LA}=20.7$~km~s$^{-1}$ respectively. It is interesting to find
that the $C_{TW}$ is much smaller than that in the graphite plane.
This effect is expected to be measured in experiments with
technical developments. The frequencies of RBM, $\vec{e}_{r}$
optical (OP) mode, $\vec{e}_{\phi}$ OP mode and $\vec{e}_{z}$ OP
mode are calculated as 538.6, 1081.5, 1544.2 and 1487.7~cm$^{-1}$
respectively. At $(\kappa, n)=\pm(\alpha, 1)$ with
$\alpha=2.01$~radian, there are two transverse acoustic modes--
i.e., the flexure modes-- with the parabolic dispersions as
$\omega^{2}=\beta^{2}(\kappa\mp\alpha)^{4}$ in the low-frequency
limits. They are the consequence of the rotational invariance of
our vibration potential.\cite{Mahan3, Jiang}

For tubes with diameter larger than approximately 0.8~nm where the
curvature effect dies away, the calculated frequencies
(Fig.~\ref{Fig:ModeVariety}), the sound velocities for various
vibration modes, and even bond length (Fig.~\ref{Fig:bondlength})
etc. are in accord with those of the tubes of ideal
geometry.\cite{Jiang} We notice that the slight difference in
force constants between the present paper and
Ref.~\onlinecite{Jiang} take no substantial effect on the lattice
vibrational properties.

Although the lattice structure for the relaxed narrow SWCNT's
deviates from that of the ideal geometry considerably, the chiral
index $(n_1,n_2)$ for the relaxed SWCNT's retains the symmetry
information inherited from the hexagonal planar lattice sheet. As
one might expect, our calculation verifies that various  physical
quantities can be expanded in terms of $\cos 3n\theta$ and $\sin
3n\theta$, $n=1,2,\ldots$, with the same expansion forms as those
of the ideal geometry, in which $\theta$ is the chiral angle
defined for the correspondent tube with ideal geometry as $
\theta=\arctan \frac{\sqrt{3}n_2}{2n_1+n_2}$. As an example, we
show our fitted polarization vectors of RBM and two optical modes
in tangential plane at $(\kappa, n)=(0,0)$ in
Table~\ref{Tab:eigenvector}. The correspondent coefficients are
shown almost the same as those in Table~IV of
Ref.~\onlinecite{Jiang}.
\section{The Young's modulus and the Poisson ratio}
The relaxed lattice configuration provides itself as the proper
minimum of the total potential energy. In particular the
equilibrium lattice configuration for the loaded SWCNT's can also
be performed, which results that the calculation for various
strain-induced physical quantities such as Young's modulus and
Poisson ratio can be straightforwardly carried out. Among the
GMV's, $h$ and $h'$ are the variables of the longitudinal degrees
of freedom, while $r$, $\alpha$ and $\alpha '$ are variables of
peripheral degrees of freedom. However, the experimentally
measured longitudinal stretching should correspond to the variable
$h$ while the variable $h'$ essentially describes the microscopic
in-cell displacement. The Young's modulus for the SWCNT's are
introduced as:
\begin{eqnarray}
Y=\left.\frac{\partial^{2} E}{\partial
\varepsilon_{11}^{2}}\right|_{\mbox{{\footnotesize zero stress
except the $(z, z)$ component}}}\;
    \label{Eq:YoungE}
\end{eqnarray}
where $\varepsilon_{11}=\frac{\Delta h}{h}$ being the longitudinal
strain, and the derivatives are taken with all the other GMV's $r,
\alpha, \alpha'$ and $h'$ being freely relaxed which corresponds
to the boundary condition that all the components of stress tensor
are zero except the $(z, z)$ component. We stress that such a
calculation of the Young's modulus for the cylindrical lattice
sheet is consistent with the classical definition of Young's
modulus in the continuum limit.\cite{Landau} Accordingly the
Poisson ratio
\begin{eqnarray}
\mu=\left|\frac{\varepsilon_{22}}{\varepsilon_{11}}\right|_{\mbox{{\footnotesize
zero stress except the $(z, z)$ component}}}\;
    \label{Eq:PoissonE}
\end{eqnarray}
with $\varepsilon_{22}$ being the peripheral strain induced under
the same condition as explained above.

The calculated Young's modulus and Poisson ratio for various
narrow SWCNT's can be well fitted as the functions of radius $r$
and chiral angle $\theta$,
\begin{eqnarray}
Y & = & 56.3-\frac{15.1+21.8\cos6\theta}{r^2} \;,
    \label{Eq:Young} \\
\mu & = & 0.166+\frac{0.235+0.246\cos6\theta}{r^{2}} \;,
    \label{Eq:Poisson}
\end{eqnarray}
where $Y$ is in the unit of eV, $r$ is in the range of
$[1.3,10]$~{\AA}, $\theta$ is the chiral angle defined for the
correspondent SWCNT with ideal geometry as explained in the
previous section, and the relative fitting error is kept less than
$1 \times 10^{-3}$. As shown in Eqs~(\ref{Eq:Young}) and
(\ref{Eq:Poisson}), the Young's modulus and Poisson ratio in large
diameter limit have the values of 56.3~eV and 0.166 respectively,
which are in good agreement with the experimental results of
graphite\cite{Weng} (see Table~\ref{Tab:YoungForGraphene}). In the
case of narrow tubes, they are significantly chirality-dependent.
For example, the Young's modulus of three SWCNT's (3,3), (4,2),
and (5,0) calculated as 57.74, 54.94, and 49.26 are considerably
different from each other, although they have almost the same
diameter.

\begin{table}[t]
\caption{The large diameter limits of Young's modulus and Poisson
ratio from our calculation and  other existing results are
compared with the experiment data.} \label{Tab:YoungForGraphene}
\begin{ruledtabular}
\begin{tabular}{cccccc}
& Experiment\cite{Weng} & Our work & Theory$^a$ & Theory$^b$ &
Theory$^c$ \\ \hline
$Y$ (eV) & 56.43 & 56.3 & 60 & 69 & 53.3\\
$\mu$ & 0.17 & 0.166 & 0.19 & 0.25 & 0.277\\
\end{tabular}
\end{ruledtabular}
$^{a}$The {\it ab initio} method;\cite{Daniel} $^{b}$The
tight-binding approximation;\cite{Rubio} $^{c}$The force constant
model\cite{Lu}
\end{table}
\begin{figure}[htpb]
    \begin{center}
        \scalebox{0.8}[0.8]{\includegraphics{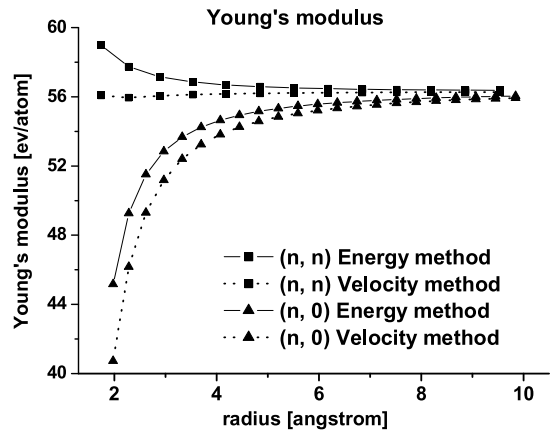}}
    \end{center}
    \caption{The Young's modulus for armchair and Zigzag tubes. The
             lines are calculated by the second derivatives of energy per
             atom with respect to the uniform strain along the tubule axis
             direction. And the dashed lines are calculated by
             Eq~(\ref{Eq:YoungV}) with velocity method.}
    \label{Fig:Young}
\end{figure}

Furthermore, we display the radius dependence of the Young's
modulus in Fig.~\ref{Fig:Young} for two species SWCNT's: armchair
tubes $(n,n)$, and zigzag tubes $(n,0)$. It shows clearly that the
Young's modulus of armchair tubes decreases with increasing of
tube diameter and always takes the values larger than that of
planar graphite. While for zigzag tubes it exhibits a contrary
behavior, i.e., increases with increasing diameter and keeps its
value always below that of graphene sheet. It was reported in
Ref.~\onlinecite{Treacy2} that among 27 samples of carbon
nanotubes, some of them are measured with the  Young's modulus
higher than that of bulk graphite. While, to our knowledge, most
of the exiting theoretical estimations\cite{Yao} can only provide
values of Young's modulus less than that of graphene sheet for any
kind of SWCNT's. Our calculation provides an evidence that
armchair tubes are stiffer while the zigzag softer due to the
chirality-dependence.
\begin{figure}[htpb]
    \begin{center}
        \scalebox{0.8}[0.8]{\includegraphics{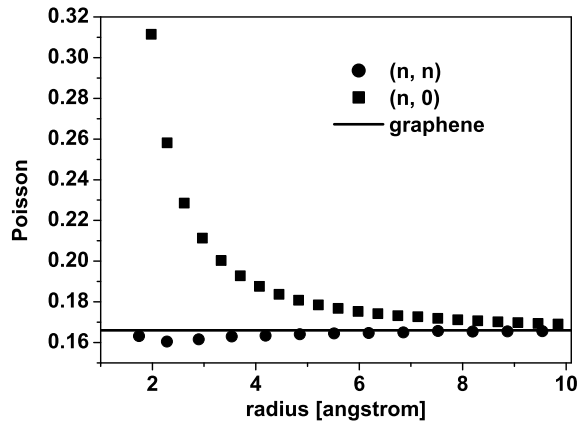}}
    \end{center}
    \caption{The Poisson ratio for different armchair and zigzag tubes.}
    \label{Fig:Poisson}
\end{figure}

It is known that following the elasticity theory of the continuum
medium, the Young's modulus as well as Poisson ratio can be
derived from the relevant sound velocities,\cite{Popov2, Landau}
\begin{eqnarray}
Y=\rho *V_{L}^{2} ,\
    \label{Eq:YoungV}
\end{eqnarray}
\begin{eqnarray}
\mu=0.5*\left(\frac{V_{L}}{V_{T}}\right)^{2}-1 ,\
    \label{Eq:PoissonV}
\end{eqnarray}
where the two expressions are also referred to the boundary
condition of only the $(z,z)$ component of stress tensor being
non-zero. Therefore, we may calculate the Young's modulus and
Poisson ratio from the relevant sound velocities obtained in the
last section shown in Fig.~\ref{Fig:Young} for comparison. We
emphasize that the Young's modulus (Poisson ratio) calculated from
the strain energy following Eqs~(\ref{Eq:YoungE}) and
~(\ref{Eq:PoissonE}) and those calculated from the sound
velocities following Eqs~(\ref{Eq:YoungV}) and
~(\ref{Eq:PoissonV}) are essentially two distinct independent
calculations. The former is in sense of a microscopic calculation
following from the total potential energy
Eqs~(\ref{Potential1})--(\ref{Potential5}), while the latter is a
sort of macroscopic calculation with the sound velocities as its
input although which is provided by the microscopic vibrational
energy (introduced in section III). We can see from
Fig.~\ref{Fig:Young}, for each of the two species of SWCNT $(n,n)$
and $(n,0)$, the Young's module produced by the two kinds of
calculations meet to each other closer and closer as the radius
increases and have the correct $r\rightarrow \infty$ limit as the
experimental Young's modulus of graphite. Such a result arises
from the intrinsic consistency between our total potential energy
expression and that of the vibrational energy. The difference
between the two kinds of calculations becomes apparent when tube
radius decreases. We understand such a phenomenon as that the
calculation following from the theory of elasticity is macroscopic
in nature which would be failure for ultra narrow tubes.

We show the calculated Poisson ratios for armchair $(n,n)$  and
zigzag $(n,0)$ tubes in Fig.~\ref{Fig:Poisson} as the function of
radius. The values of the Poisson ratio are in an order as that
those of zigzag are the largest and the armchair the smallest.
This is consistent with the correspondent result of {\it ab
initio} calculation.\cite{Daniel}
\section{conclusions}
In this paper, we proposed a lattice dynamic treatment for the
total potential energy for SWCNT's which is, apart from a
parameter for the non-linear effects, extracted from the
vibrational energy of the planar graphene sheet. Based upon the
proposal, we investigated systematically the relaxed lattice
configuration for narrow SWCNT's, the strain energy, the Young's
modulus and Poisson ratio, as well as the lattice vibrational
properties respected to the relaxed equilibrium tubule structure
with the same set of force constants independent of tube
structures. In particular, with the application of GMV's we not
only successfully extend the locally introduced-total potential
energy to the whole tube, but also make the lattice dynamics for
SWCNT's with relaxed geometry becomes treatable. Comparing with
the corresponding results obtained from the tubes with ideal
geometry, we verified that the relaxation effect brings the bond
length longer and the frequencies of various optical vibrational
modes softer. Moreover, our approach not only deduces the proper
large diameter limit values of Young's modulus and Poisson ratio,
but also provides a strong evidence that SWCNT's with different
chirality could be stiffer or softer than the planar graphene
sheet. The calculated RBM and optical modes in tubule tangential
plane are also in good agreement with the experimental
measurements. In addition, we realized  that the chiral
symmetry-based general expansion formulae for different physical
quantities survive for SWCNT's with relaxed geometry in the sense
to replace the chiral angle by its counterpart defined in tubes of
ideal geometry.

\end{document}